\let\myover=\over   
\documentclass[12pt,a4paper]{article}
\usepackage{amssymb,amsmath,epsfig}
\def\be{\begin{equation}}
\def\ee{\end{equation}}

\def\l{\left(}
\def\r{\right)}

\newcommand{\bg}{\begin{gather}}
\newcommand{\eg}{\end{gather}}

\def\be{\begin{equation}}
\def\ee{\end{equation}}

\def\half{\frac{1}{2}}

\begin{document}
\let\over=\myover  
\def\half{{1 \over 2}} 
\title{On Electroweak Baryogenesis in Gauge Mediated Models with
Messenger-Matter Mixing}
\author{ D.~S.~Gorbunov\thanks{{\bf e-mail}: gorby@ms2.inr.ac.ru}, \\
{\small{\em
Institute for Nuclear Research of the Russian Academy of Science, }}\\
{\small{\em
60th October Anniversary prospect 7a, Moscow 117312, Russia
}}} 
\date{}
\maketitle
\begin{abstract} 
  \small We consider feasibility of electroweak baryogenesis in 
gauge mediated supersymmetry breaking models with messenger-matter
mixing. We present a class of models where electroweak baryogenesis
produces sufficient amount of baryon asymmetry. The main 
features of these models are ({\it i}) large mixing between messengers
and right stop and ({\it ii}) fairly narrow region of viable $\tan\beta$. 
\end{abstract} 

\section{Introduction}
Among various puzzles of Nature the problem of baryogenesis is one
of the most intriguing. Since pioneering 
works~\cite{sakharov}, necessary conditions for the generation of 
the baryon asymmetry are well-known. These are baryon number
violation, C- and CP-nonconservation and departure from thermal
equilibrium at a certain stage of the evolution of the Universe. 
Search for reliable mechanisms of baryogenesis is limited to 
theories where these conditions are met. 

One of the most appealing mechanisms is electroweak 
baryogenesis~\cite{ew,EB-reviews}. It
was proposed that this mechanism works 
during the electroweak phase transition. Unlike other mechanisms,
electroweak baryogenesis was thought to be inherent in the Standard
Model (SM) and
seemed to require no additional fields. However, the baryon
asymmetry tends to be washed out by anomalous electroweak processes if
the latter come into thermal equilibrium after the phase transition 
completes. To prevent this wash-out, the electroweak phase 
transition has to be of the (strong enough) first order; this 
imposes a constraint on the expectation value of Higgs field $\upsilon (T)$ 
(with $\upsilon (0)\approx 245$~GeV) at the critical 
temperature $T_c$~\cite{shaposhnikov}, 
\begin{equation}
{\upsilon(T_c)\over T_c}\gtrsim 1\;.
\label{constraint}
\end{equation}
In the case of SM this
requirement implies the bound on the Higgs boson mass~\cite{EB-reviews},
\begin{equation}
m_h<40~{\rm GeV}\;,
\label{h-SM}
\end{equation}  
which is inconsistent with present
experimental limits~\cite{Higgs-limit}. 
Hence electroweak baryogenesis fails in the Standard
Model.~\footnote{There are recent attempts (see, e.g., Ref.~\cite{dima}) to
revive electroweak baryogenesis in SM by invoking the dynamics of
preheating.} 

Similar situation takes place in the Minimal Supersymmetric Standard
Model (MSSM). Although the corresponding bounds on the parameters of
MSSM are significantly weaker, almost entire parameter space consistent
with successful  electroweak baryogenesis is excluded by existing
experimental data. Still, 
electroweak baryogenesis works in 
MSSM with light Higgs, light $\tilde{t}_R$, 
heavy $\tilde{t}_L$ , small $\tilde{t}_R-\tilde{t}_L$ 
mixing, and charginos with fairly degenerate
masses~\cite{light-stop} (for brief reviews,
see Refs.~\cite{9704347,9903274}). This spectrum was discussed in the framework
of models with gravity mediation of 
supersymmetry breaking~\cite{light-stop}, and it was observed that 
the price for this ``light stop''
solution is non-universal boundary conditions for
soft terms at GUT scale. 

Recently, the phenomenology of Gauge Mediated
Supersymmetry Breaking (GMSB) models attracted considerable attention
(for a review, see Ref.~\cite{revGMM}). In these models, 
supersymmetry breaking occurs in a separate sector. 
Supersymmetry breaking terms 
in the visible sector are generated due to special fields
(messengers) charged under SM gauge group. The
soft masses of superpartners 
are determined by their quantum numbers and 
are proportional to the corresponding gauge coupling constants. Consequently, 
all squarks are very heavy and electroweak baryogenesis seems
not to work. So, it has been argued that 
the favorite mechanism to produce the baryon asymmetry in GMSB models
is the Affleck--Dine
baryogenesis~\cite{AD-baryogenesis}. 

This letter addresses the possibility of electroweak
baryogenesis in GMSB with messenger-matter mixing. This mixing is
natural since messengers carry the same quantum numbers as ordinary
fields in MSSM~\cite{dns-mixing}. The constraints on mixing parameters
imposed by the absence of
lepton flavor violating processes and FCNC are not particularly 
strong~\cite{we-1,we-2}. Moreover, 
it was pointed
out that some scalar masses may be significantly reduced at large
values of mixing terms 
without any contradiction to experimental limits~\cite{we-2}. We will
see that the small mass of stop required for successful electroweak
baryogenesis may be explained in this way. 
We will show also that all other conditions on the spectrum may be 
satisfied. 

\section{Light stop window for electroweak baryogenesis}

In any model of baryogenesis two
main questions arise: {\it i}) What is the mechanism of the generation
of the baryon asymmetry? {\it ii}) What protects the 
baryon asymmetry from being washed out by sphaleron processes?

In the framework of ``light stop'' solution, the answers to these
questions are as follows. 

{\it i}) The relevant source of CP-violation is the phase $\phi_{CP}$
of $\mu$-term in Higgs superpotential. 
Baryogenesis is fueled by CP-asymmetry in chargino flux
through expanding bubble wall. 
The realistic amount of baryon asymmetry $n_B/s$ is
   produced by sphaleron processes provided that chargino and neutralino
masses are
not much larger than
   the critical temperature $T_c\sim$100~GeV. The
   resulting asymmetry is determined by the ratios of masses to the
critical temperature,
   and degenerate $\mu$ and wino mass $M_2$ are favorable. Though the
general tendencies are clear, the actual calculated values 
of $n_B/s$ depend on the approximations made. In what follows we
make use of the constraints on the values of CP-phase and
   the level of degeneracy presented in Ref.~\cite{9907460}. The
smallest allowed value of $\phi_{CP}$ is $|\phi_{CP}|=10^{-4}$; at
this value one requires $\mu=M_2$. 
In the opposite case of large CP-violating phase, $|\sin\phi_{CP}|=1$,
the allowed region of masses is $\mu=\epsilon M_2$, 
$0.4\lesssim\epsilon\lesssim 2.5$ at $\mu$,$M_2$$\simeq T_c$ and
   $0.5\lesssim\epsilon\lesssim 2$ at $\mu$,$M_2$$\simeq
   3T_c$. The favorable region of the mass of CP-odd Higgs boson is 
$m_A\lesssim 300$~GeV, otherwise $n_B/s$ is suppressed
   by $m_A^{-2}$~\cite{9702409,9704347}. For heavier CP-odd Higgs, the
   values of $\mu$
   and $M_2$ are to be more degenerate in order that the 
baryon asymmetry at given $\phi_{CP}$ be the same as at small $m_A$. 

{\it ii}) As mentioned above, electroweak phase
   transition has to be of the first order, so that the
   constraint~(\ref{constraint}) is satisfied. Light
   stop helps in strengthening the phase transition. The
   largest baryon asymmetry is obtained in the context of MSSM
   with right stop plasma mass vanishing at the critical
   temperature, when the condition~(\ref{constraint}) becomes~\cite{AA}  
\begin{equation}
1\lesssim{\upsilon(T_c)\over T_c}=\l{\upsilon(T_c)\over T_c}\r_{SM}+{2 m_t^3\l 
1-{\tilde{A}_t^2\over m_Q^2}\r^{3/2}\over \pi\upsilon m_h^2}\;,
\label{MSSM-constraint}
\end{equation}  
where $\tilde{A}_t=|A_t-\mu\cot\beta|$, and  $A_t$ is the stop trilinear
soft term, $m_Q$ is the left stop mass and $m_t$ is the on-shell running
top quark mass in the $\overline{MS}$ scheme. Hereafter $m_h$ denotes
the mass of the lightest Higgs boson. We
consider the case $m_A\gg T_c$, which is relevant to GMSB models; in
this case CP-odd Higgs does not affect the electroweak phase transition. 
The first term on the right hand side of 
Eq.~(\ref{MSSM-constraint}) is the ordinary Standard Model
contribution, 
$$
\l{\upsilon(T)\over T}\r_{SM}\simeq \l{40~{\rm GeV}\over m_h}\r^2\;.
$$  
The second term in Eq.~(\ref{MSSM-constraint}) is the contribution
from light right stop. 
Current limits on the lightest Higgs mass~\cite{Higgs-limit}, 
$m_h\gtrsim100$~GeV, combined with the
inequality~(\ref{MSSM-constraint}) impose a constraint 
on left-right mixing 
\begin{equation}
\tilde{A}_t/m_Q\lesssim0.5\;.
\label{R*}
\end{equation}
At larger $m_h$ left-right 
mixing has to be smaller. At $\tilde{A}_t=0$
Eq.~(\ref{MSSM-constraint}) gives the upper bound on the lightest
Higgs boson mass in the theory with successful electroweak baryogenesis,
$m_h\simeq 115$~GeV~\cite{9903274}. It is worth noting 
that the result~(\ref{MSSM-constraint}) has been obtained by
making use of improved one-loop effective potential. Higher order
corrections make the phase transition slightly 
stronger (for a brief review and references, see Ref.~\cite{9903274}) 
and will be neglected in what follows.  

Zero right stop plasma mass at the critical
   temperature implies 
\begin{equation}
m_{\tilde{t}_R}^{2(eff)}=m_U^2+\Pi_R(T_c)\approx0\;,
\label{mt-zero}
\end{equation}
where $m_U$ is the low energy value of the right stop soft mass term
and $\Pi_R(T_c)$ is the finite temperature contribution to the effective
squared mass
which is of order $T_c^2$~\cite{light-stop}. Hence one needs negative $m_U^2\sim
-(100$~GeV)$^2$, which in principle may result in the existence of charge-
and color-breaking (local) vacuum. A conservative requirement is that the
physical vacuum has lower energy than the color-breaking minimum. The
latter condition at, as an example, $\tilde{A}_t=0$ yields 
approximately~\cite{light-stop} 
\begin{equation}
|m_U|<m_{crit}=\l{m_h^2\upsilon^2\alpha_3\pi\over 3}\r^{1/4}\;,
\label{min-bound}
\end{equation}
that is $|m_U|\lesssim 95$~GeV for $m_h=100$~GeV. One can relax this constraint
by considering the physical vacuum as a metastable but long-living 
minimum. However, the inequality 
\begin{equation}
m_U^2+\Pi_R(T_c)>0
\label{temp-bound}
\end{equation}
has to be satisfied in any case, otherwise the 
Universe would be driven to a charge- and
color-breaking minimum at $T>T_c$ (for a discussion see
Ref.~\cite{light-stop}).  

Let us collect the requirements which are imposed on the 
theory with ``light stop'' solution: 
\begin{enumerate}
\item Right-left mixing in stop sector is small, $\tilde{A}_t<0.5m_Q$
at $m_h>100$~GeV, and the heavier the lightest Higgs boson, the
smaller the mixing. 
\item At $m_h>115$~GeV there is no window for
electroweak baryogenesis. Consequently, the mass of the lightest Higgs
boson belongs to the interval 100~GeV$<m_h<$115~GeV, where
the lower bound comes from experiment.   
\item The favorable interval of the mass of CP-odd Higgs boson is 
$150~{\rm GeV}\lesssim m_A\lesssim 300~{\rm GeV}$. At smaller $m_A$
the phase transition is weaker, while at larger $m_A$ the baryon
asymmetry is suppressed by $m_A^{-2}$. 
\item The soft mass squared of right stop is
negative and of order $-T_c^2$. Inequalities~(\ref{min-bound}),
(\ref{temp-bound}) (or similar ones at non-zero $\tilde{A}_t$) have to
be satisfied.  
\end{enumerate}

\section{Gauge Mediated Models with messenger-matter mixing}

Let us recall the spectrum of superpartners in GMSB models~\cite{revGMM}. In 
simple versions, messengers belong to complete vector-like 
GUT (e.g., $SU(5)$) multiplets. The induced soft terms depend crucially on 
$n=n_5+3n_{10}$, with $n_5$ and $n_{10}$ being the numbers of $5+\bar{5}$
and $10+\bar{10}$ messenger generations. Indeed, the spectrum of 
superpartners in these models at the messenger scale $M_m$ is 
\begin{equation}
\label{gaugin}
M_{i} = nc_i\frac{\alpha_i}{4\pi}\Lambda
f_1({\Lambda\over M_m})\;,
\end{equation}
\begin{equation}
\label{scalmas}
m^2=2n\Lambda^2\sum_{i=1}^3C_ic_i\l\frac{\alpha_i}{4\pi}\r^2
f_2({\Lambda\over M_m})\;,
\end{equation}
where $M_{i}$ denote gaugino masses and $m$ are soft masses of
the scalar superpartners of fermions of the Standard Model. 
Here $\alpha_1=\alpha/\cos^2\theta_W$, 
$\alpha_2=\alpha/\sin^2\theta_W$, $\alpha_3$ are
gauge coupling constants of electroweak and strong interactions; $c_1=5/3$,
 $c_2=c_3=1$; $C_3=4/3$ for color triplets (zero for singlets),
 $C_2=3/4$ for weak doublets (zero for singlets),
$C_1=\l\frac{Y}{2}\r^2$, where $Y=2(Q_{em}-T_3)$ is the weak
hypercharge, and $\Lambda<M_m$ is a dimensional
parameter related to the scale of supersymmetry breaking. 
The functions $f_1$ and $f_2$ weakly depend on their argument and are close to
1 in the most part of the domain of definition. We consider soft
mixing term in the Higgs sector $B_\mu$
(or $\tan\beta$) as a free parameter. The value of $\mu$ is 
determined by the relation  
\begin{equation}
\mu^2=-\half M_Z^2+{m_{h_D}^2-m_{h_U}^2\tan^2\beta\over\tan^2\beta-1}\;, 
\label{electroweak-consistency}
\end{equation}
where $m_{h_U}$ and $m_{h_D}$ are soft masses of up- and down-Higgs
bosons. Note that 
gaugino masses grow as $n$, whereas scalar ones grow as $\sqrt{n}$. 
It is this behavior that enables one to obtain the 
degeneracy $\mu\sim M_2$ favorable for successful generation of baryon
asymmetry.  

Electroweak symmetry breaking leads to additional contributions 
to the soft mass spectrum. 
D-terms drive the masses up while mixing between left and right
scalars increases the splitting between their masses. 
Taking into account these corrections and 
the renormalization 
group evolution from $M_m$ to $M_{SUSY}$ one calculates the low
energy spectrum. In fact, reasonable estimates for the low energy
soft masses are obtained by plugging $\alpha_i=\alpha_i(M_{SUSY})$ into
Eqs.~(\ref{gaugin}) and (\ref{scalmas}). 
As concerns the trilinear soft terms, their values at $M_m$ 
are additionally suppressed by coupling constants in comparison
with Eqs.~(\ref{gaugin}), (\ref{scalmas}) and become significant 
at low energies only due to the renormalization group evolution. The 
largest low energy trilinear coupling is $A_t$, which is numerically
$A_t\simeq M_2$.  

We consider messengers which are odd under R-parity. Then the
components of the fundamental messengers have the same quantum numbers
as left leptons and right down-quarks, while components of
antisymmetric messengers have quantum numbers of right leptons, left
quarks and right up-quarks. Therefore, symmetries allow for mixing
between messengers and matter
fields~\cite{dns-mixing}. 
In the case of one fundamental messenger multiplet, the mixing terms in
the superpotential are 
\begin{equation}
{\cal W}^{(5)}_{mm}=H_DL_mY_i^{(5)}E_i+H_DD_mX_i^{(5)}Q_i\;,
\label{sup-pot-mixing}
\end{equation}
where $i=1,2,3$ counts matter generations and subscript $m$ refers to 
messenger fields. We use the standard notations for MSSM superfields 
($E_i$ are right leptons and $Q_i$ are left
up-quarks), $Y_i^{(5)}$ and $X_i^{(5)}$ are mixing parameters, 
$L_m$ and $D_m$ are messenger superfields with quantum numbers of left
leptons and right down-quarks, respectively. 

When messengers are integrated out, mass matrices of scalar
fields acquire negative contributions,
$$
\delta m_{ij}^2\sim -10^{-2}\Lambda^2Y^*_iY_j\;,
$$
that emerge from one-loop diagrams with messenger fields running in 
loops~~\cite{dns-mixing} (here $Y_i$ generically stands for either
$Y_i^{(5)}$ or $X_i^{(5)}$). 
In principle, these terms give rise to flavor violating processes 
($\mu\to e\gamma$, $b\to s\gamma$, etc.); the corresponding limits on
$Y_i$ are derived in Ref.~\cite{we-2}. It is important for our
discussion that the mixing terms also reduce 
some scalar masses. Indeed, let us consider colorless sector and 
neglect left-right mixing. 
Then the eigenvalues of right slepton mass matrix are 
$$
\{m_R^2\;,~~m_R^2\;,~~m_R^2-\sum_{i=1}^3\delta m_{ii}^2\}\;,
$$
where $m_R$ is the soft mass of right slepton. The squark masses are
modified in a similar way. This hints towards the possibility to have
the right stop soft mass 
term $m_{\tilde{t}_R}^2$ negative, as required by the ``light stop'' scenario
of electroweak baryogenesis. However, fundamental messengers are not
suitable for this purpose, as they do not mix with right stop. Hence,
messenger fields belonging to antisymmetric representation are to be
involved. 

\section{An example of GMSB model with ``light stop'' window}

As a concrete example of a model where all criteria of 
``light stop'' solution are satisfied, we consider GMSB
model with $n_{10}$ antisymmetric messengers and non-zero 
coupling $Y_3$ just
between right stop $U_3$ and corresponding 
messengers $Q_m$, 
\begin{equation}
{\cal W}^{(10)}_{mm}=Y_3H_UU_3Q_m\;.
\label{one-loop}
\end{equation}
When messengers are integrated out, no additional mixing appears in
squark matrix, 
so there is no problem with FCNC~\footnote{In
what follows we will require rather large value of $Y_3$. A 
model with three large mixing terms of the same order of magnitude 
in right up-squark sector is ruled out by the absence of
FCNC. On the other hand, if we assume the hierarchy 
between messenger-matter couplings similar to the hierarchy between SM
fermion Yukawa couplings, we obtain a ``natural'' model with large
messenger--stop mixing and suppressed FCNC.}. As a result of
mixing~(\ref{one-loop}), the right stop mass squared,
$m_{\tilde{t}_R}^2$, and up-Higgs mass squared, $m_{h_U}^2$, get negative 
contributions~\cite{we-2}, 
\begin{equation}
\delta m_{\tilde{t}_R}^2\approx-n{\Lambda^2\over
8\pi^2}|Y_3|^2\;f_3\!\l{\Lambda\over M_m}\r\;,~~~~~
\delta m_{h_U}^2\approx-n{3\Lambda^2\over 16\pi^2}|Y_3|^2
\;f_3\!\l{\Lambda\over M_m}\r
={3\over 2}\delta m_{\tilde{t}_R}^2\;,
\label{mass-cont}
\end{equation}
where $f_3(\Lambda/M_m)=(1/6)(\Lambda/M_m)^2$ at $\Lambda/M_m$ not
very close to 1. 
These terms come from one-loop diagrams involving the Yukawa 
interaction~(\ref{one-loop}), with messengers running in loops. 

For given $\Lambda$ and $n=3n_{10}$ it is
possible to tune $Y_3$ and obtain a negative value of $m_{\tilde{t}_R}^2$ of
order of $-m_{crit}^2$. In particular, Eq.~(\ref{mt-zero}) is obeyed
at 
\begin{equation}
|Y_3|^2\simeq8\pi^2{m_{\tilde{t}_R}^2+m_{crit}^2\over 
3n_{10}\Lambda^2}\;f^{-1}_3\!\l{\Lambda\over M_m}\r\;.
\label{Y}
\end{equation}
where $m_{\tilde{t}_R}^2$ as given by Eq.~(\ref{scalmas}) is typically
much larger than $|m_{crit}|^2$. The value of the messenger-stop
Yukawa coupling turns out to be fairly large. As an example, at
$\Lambda/M_m=0.5$ one has $Y_3\simeq 4\sqrt{2}\alpha_3\simeq 0.6$. 
In this range of $Y_3$, Eq.~(\ref{min-bound}) is also straightforward
to fulfill. Hence, the fourth
requirement of section 2 may be satisfied by tuning $Y_3$. 

We turn to the other three requirements. The lower bound on the 
lightest Higgs boson mass is experimental and has to be satisfied 
regardless of baryogenesis. This is not the only constraint on the
parameter space coming from collider experiments: as there is no
evidence for light right sleptons, small
values $\Lambda\sqrt{n}<30$~TeV are forbidden. At the level of our
accuracy we approximate the one-loop enhanced mass of the 
lightest Higgs boson as follows, 
\begin{equation}
m_h^2=M_z^2\cos^22\beta+{3\sqrt{2}\over2\pi^2}G_Fm_t^4\l\log\l{m_Q^2\over
m_t^2}\r +{\tilde{A}_t^2\over m_Q^2}\r\;.
\label{higgs-mass}
\end{equation}  
Since one of the requirements of the ``light stop'' solution is small
left-right mixing in stop sector (see Eq.~(\ref{R*})),
equation~(\ref{higgs-mass}) implies the lower bound on $\tan\beta$
depending on the key GMSB parameter, $\Lambda\sqrt{n}$, that
determmines the value of $m_Q$ through Eq.~(\ref{scalmas}). One
obtains $\tan\beta\gtrsim3.5$ at $\Lambda\sqrt{n}=30$~TeV, 
$\tan\beta\gtrsim 2.5$ at $\Lambda\sqrt{n}=50$~TeV and 
$\tan\beta\gtrsim 1.5$ at $\Lambda\sqrt{n}=100$~TeV. Within our
approximation, the largest value of $m_h$ consistent with
``light stop'' solution, $m_h<115$~GeV, is achieved at very large 
$\Lambda\sqrt{n}$ and hence $m_Q$: $\Lambda\sqrt{n}\gtrsim400$~TeV, 
$m_Q\gtrsim4$~TeV. The reason is that, as we will see below,
successful baryogenesis requires rather small values of $\tan\beta$. 
It worth noting that similar constraints are imposed on the parameters of
SUGRA models with ``light stop'' solution. 
 
Let us
proceed with other requirements. 
Before discussing concrete cases of small and large CP-violating phase
$\phi_{CP}$, let us make a few general remarks. 

First, 
mixing~(\ref{one-loop}) drives $m_{h_U}^2$
deep into the region of negative values, 
\begin{equation}
m_{h_U}^2=m_{h_D}^2-{3\over8\pi^2}m_Q^2\l\log{M_m\over
m_Q}+{3\over 2}\r+{3\over 2}\delta m_{\tilde{t}_R}^2\;,
\label{higgs-contribution}
\end{equation}     
that leads to very large
$\mu$ because of Eq.~(\ref{electroweak-consistency}). This implies
that the CP-phase has to be relatively large, 
since $m_A\sim\mu$, and the baryon asymmetry is suppressed by
the mass of heavy CP-odd Higgs boson. Moreover, since an approximate
degeneracy $\mu\simeq M_2$ is required, wino is also heavy. 
This makes a potential problem, because in GMSB models with complete messenger
multiplets one has $A_t\simeq M_2$, and there appears large left-right mixing.
The cure is to make $\tilde{A}_t=|A_t-\mu\cot\beta|$ small,
$\tilde{A}_t<0.5 m_Q$, by choosing suitable sign of $\mu$ and 
tuning  $\tan\beta$. 

Second, the degeneracy between $\mu$ and $M_2$ will be achieved by a
suitable choice of the number of messengers, $n_{10}$. Indeed, 
these two mass parameters scale differently with the 
number of messenger fields. Namely, from 
Eq.~(\ref{electroweak-consistency}) one finds that $\mu$ is determined
by scalar soft masses, so it grows as $\sqrt{n}$ (see
Eq.~(\ref{scalmas})), while $M_2$ grows as $n$ (see
Eq.~(\ref{gaugin})). 

One can estimate viable values of parameters of this model 
by the following simple chain of reasonings. 
Squarks are the heaviest scalars in this theory,
because the hierarchy of soft terms are governed by corresponding gauge
couplings. Since $\Lambda\sqrt{n}\gtrsim30$~TeV, the model has heavy strong
sector, $m_Q^2\gg m_Z^2$. The 
``light stop'' solution requires $|m_Q^2+\delta
m_{\tilde{t}_R}^2|\sim M_z^2$, so $|\delta m_{\tilde{t}_R}^2|$ is also
large, and $m_{h_U}^2\approx{3\over 2}\delta m_{\tilde{t}_R}^2$ 
(see Eqs.~(\ref{higgs-contribution}), (\ref{mass-cont})). Then the 
relation~(\ref{electroweak-consistency}) may be approximated as
follows, 
\begin{equation}
\mu^2={3\over 2}m_Q^2{\tan^2\beta\over \tan^2\beta-1}\;.
\label{mu-approx}
\end{equation} 
The ``light stop'' solution requires almost degenerate spectrum, 
$\mu=\epsilon M_2$ with $1/2<\epsilon<2$. In view of the relation 
$A_t\simeq M_2$, Eq.~(\ref{R*}) implies 
\begin{equation}
\left|{1\over\epsilon}-{1\over\tan\beta}\right|\lesssim\half{m_Q\over\mu}\;,
\label{mix-approx}
\end{equation} 
Making use of Eq.~(\ref{mu-approx}) we obtain from the 
inequality~(\ref{mix-approx}) 
\begin{equation}
\left|{1\over\epsilon}-{1\over\tan\beta}\right|\lesssim{1\over\sqrt{6}}
\sqrt{1-{1\over\tan^2\beta}}\;. 
\label{model-limit}
\end{equation}  
This determines the viable values of 
$\tan\beta$ for given degeneracy $\epsilon$. At
$\epsilon=1$ ($M_2=\mu$) one has $\tan\beta\lesssim 1.5$, at 
$\epsilon=1.5$ one has $\tan\beta\lesssim 4$ while at
$\epsilon=2$ one obtains $\tan\beta\lesssim 10$. 

By making use of Eq.~(\ref{mu-approx}) and Eqs.~(\ref{gaugin}),
(\ref{scalmas}) we find the number of
messengers required for succesfull electroweak baryogenesis at given
$\epsilon$ and $\tan\beta$,
\begin{equation}
n_{10}\simeq{4\over
3\epsilon^2}{\alpha_3^2\over\alpha_2^2}{\tan^2\beta\over\tan^2\beta-1}\;. 
\label{n-approx}
\end{equation}
The estimates are: 
$n_{10}\gtrsim24$ at $\epsilon=1$ and $\tan\beta\lesssim1.5$, 
$n_{10}\gtrsim6$ at $\epsilon=1.5$ and $\tan\beta\lesssim4$, 
$n_{10}\gtrsim3$ at $\epsilon=2$ and large $\tan\beta\lesssim 10$. 

Until now we discussed mostly the bounds coming from the survival of
baryon asymmetry after the electroweak phase transition. Additional
bounds come from the calculations of the
baryon asymmetry along the lines of Ref.~\cite{9907460}. Let us
present the results for the cases of small and large $\phi_{CP}$. They
depend on the value of $\Lambda$ which determines the mass of the
CP-odd Higgs boson $m_A$, and hence the suppression factor in the
baryon asymmetry.

The smallest CP-phase corresponds to the case of complete degeneracy
between $M_2$ and $\mu$, i.e., $\mu/M_2\equiv\epsilon=1$. In this case
$\tan\beta\lesssim 1.5$, so one has to take 
$\Lambda\sqrt{n}\gtrsim 100$~TeV in order to obtain $m_h>100$~GeV. 
As discussed above, the number of messengers is large, $n_{10}\gtrsim
24$. This results in 
the large value of Higgsino mass $\mu\gtrsim 2200$~GeV and 
large $m_A\approx\mu/\sin\beta$, that 
makes the phase
transition stronger, but reduces the amount of baryon asymmetry at
given $\phi_{CP}$. The realistic value of $n_B/s$ is obtained 
at $\Lambda\sqrt{n}=100$~TeV with $|\sin\phi_{CP}|\gtrsim0.008$. 
At $\Lambda\sqrt{n}=200$~TeV we find 
$\tan\beta\lesssim 1.5$, $n_{10}\gtrsim 24$, $\mu\gtrsim 4400$~GeV and 
$|\sin\phi_{CP}|\gtrsim0.03$. 

In models with large
CP-violation phase ($|\sin\phi_{CP}|=1$), fine tuning between $M_2$ and
$\mu$ may be somewhat relaxed, and the constraints on $n_{10}$ and 
$\tan\beta$ are, of course, weaker. As an example, let us consider the
case $\mu=1.5M_2$. At $\Lambda\sqrt{n}=50$~TeV 
one obtains $n_{10}\gtrsim 6$, 
$\mu\gtrsim$ 800~GeV and $\tan\beta \lesssim 4$, 
while at $\Lambda\sqrt{n}=100$~TeV one
gets $n_{10}\gtrsim 6$, $\mu\gtrsim$ 1600~GeV and $\tan\beta \lesssim
4$. 

Let us note in passing that the case $\mu<M_2$ is not favorable 
for the ``light stop''
solution. Indeed, it follows from Eq.~(\ref{model-limit}) that
the viable region of $\epsilon\equiv\mu/M_2<1$ is very narrow and 
$\tan\beta$ is smaller than $1.5$ at $\epsilon$ within this interval. 
Also, small values of $\Lambda\sqrt{n}$ are disfavored. As an example,
at $\Lambda\sqrt{n}=30$~TeV one finds $\mu\gtrsim$ 450~GeV, and the
absence of right slepton~\footnote{In the case of large $\mu$ which is
of relevance throughout our discussion the masses squared of right
sleptons acquire negative contributions due to electroweak symmetry
breaking, which are proportional to $\mu\tan\beta$~\cite{revGMM}.}
with mass smaller than 45~GeV requires 
$\tan\beta\lesssim 3$. This is in contradiction with the bound coming
from the lightest Higgs boson mass, $\tan\beta\gtrsim 3.5$.

One concludes that all requirements for the ``light stop'' solution are
satisfied in this model, hence electroweak baryogenesis
is capable of generating sufficient amount of baryon asymmetry in GMSB models. 
The region of viable parameters is fairly
narrow due to strong restrictions coming from the search for the
lightest Higgs boson. Indeed, $\tan\beta$ in models with ``light
stop'' solution tends to be small, while existing limits on the
lightest Higgs boson mass favor high $\tan\beta$. 
At small CP-phase the suitable region stretches along
$\tan\beta\sim1.5$ and the smallest CP-phase, $\phi_{CP}\sim10^{-2}$, 
is possible at $\Lambda\sqrt{n}\simeq$100~TeV. At
$\phi_{CP}\sim 1$ this region becomes
wider and extends to $\tan\beta\simeq 4$. The suitable region of 
parameter space is the largest at the smallest possible $m_h$. 
A drawback of the  
model is a large number of messenger fields, 
so the gauge coupling constants become large below the scale of
possible Grand Unification. 

Electroweak baryogenesis would work better in a model where left
squarks are heavy, whereas right squarks are light 
(or $\mu$ is relatively small) and $A_t$ is small. In the framework of
GMSB these features appear in models with 
additional large mixing terms or extended messenger content. 
Let us briefly discuss these possibilities. 

One can introduce additional mixing (e.g., between d-squark
and corresponding messengers), which provides large negative
contribution to mass squared of down-Higgs. Then larger 
values of $\tan\beta$ may become viable. Likewise, the value of 
$\mu$ may be smaller,
especially at low $\tan\beta$, which would extend the region of
suitable values of CP-phase.  

One can also consider messengers which do not form complete GUT
multiplets.~\footnote{The same
observations apply to models where messengers carrying different
quantum numbers are characterized by different scales $\Lambda$.} 
Let us denote the effective numbers of messengers generating 
gluino mass and wino mass as $n_c$ and $n_w$, respectively. 
Then the relevant quantity will be $n_w^2/n_c$ instead of
$n_{10}$. Therefore, it will be possible to reduce
significantly the number of messenger fields and relax the problem
of unification of gauge coupling constants in a theory with ``light stop''
window.    

There is yet another possibility related to messengers belonging
to incomplete GUT multiplets. We outline it by making use of the
simplest example
of only lepton-like messengers without any messenger-matter mixing. In
this model, 
right stop acquires negative mass squared due to
renormalization group evolution, in analogy to the up-Higgs in MSSM. There
is a wide region of parameter space, where electroweak symmetry
breaking occurs but left squarks remain heavy. In this model the 
favorable hierarchy 
$$
m_Q\gg m_{\tilde{t}_R}\;,A_t
$$
appears naturally. The reason for the hierarchy between $m_Q$ and $A_t$ 
is that the largest contribution to the low energy value of
$A_t$ comes from two-loop diagrams involving gluino; in this model
gluino is light, that reduces $A_t$ significantly. 
As a consequence, a fairly wide region of $\tan\beta$ will be
consistent with the ``light stop'' solution in this model. 

\section{Conclusions}

We have demonstrated that electroweak baryogenesis may produce 
acceptable amount of baryon asymmetry in the framework of GMSB models
with messenger-matter mixing. The required MSSM spectrum is similar to
the so-called ``light stop window'' of SUGRA models~\cite{light-stop}. 
We have plainly mapped that ``window'' onto the space of parameters of
GMSB models. We have found that severe constraints are to be imposed
on the GMSB parameters. At
least one of the mixing terms has to be quite large in order to make 
right stop lighter than top. In the explicit 
example presented in this letter, 
the minimal possible value of CP-phase is
reached at a large number of antisymmetric messenger generations,
$n_{10}\gtrsim 24$. With maximal CP-violation ($\phi_{CP}\sim 1$) one has 
$n_{10}\gtrsim 6$. Another property of GMSB models with electroweak
baryogenesis is $\tan\beta\lesssim4$. 
We have briefly outlined also extensions of this model which
have better properties with respect to 
electroweak baryogenesis. Let us
note finally, that additional sources of CP-violation, which may
originate from messenger-matter mixing, may enhance the electroweak
production of baryon asymmetry. 

\section{Acknowledgments} The author is 
indebted to V.~Rubakov and S.~Dubovsky 
for useful discussions. This work was
supported in part under Russian Foundation for Basic Research grant
99-02-18410 and by the Russian Academy of Science, JRP grant \# 37.   

\def\ijmp#1#2#3{{\it Int. Jour. Mod. Phys. }{\bf #1~} #3 (19#2)}
\def\pl#1#2#3{{\it Phys. Lett. }{\bf B#1~} #3 (19#2)}
\def\zp#1#2#3{{\it Z. Phys. }{\bf C#1~} #3 (19#2)}
\def\prl#1#2#3{{\it Phys. Rev. Lett. }{\bf #1~} #3 (19#2)}
\def\rmp#1#2#3{{\it Rev. Mod. Phys. }{\bf #1~} #3 (19#2)}
\def\prep#1#2#3{{\it Phys. Rep. }{\bf #1~} #3 (19#2)}
\def\pr#1#2#3{{\it Phys. Rev. }{\bf D#1~} #3 (19#2)}
\def\np#1#2#3{{\it Nucl. Phys. }{\bf B#1~} #3 (19#2)}
\def\mpl#1#2#3{{\it Mod. Phys. Lett. }{\bf #1~} #3 (19#2)}
\def\arnps#1#2#3{{\it Annu. Rev. Nucl. Part. Sci. }{\bf #1~} #3 (19#2)}
\def\sjnp#1#2#3{{\it Sov. J. Nucl. Phys. }{\bf #1~} #3 (19#2)}
\def\jetp#1#2#3{{\it JETP Lett. }{\bf #1~} #3 (19#2)}
\def\app#1#2#3{{\it Acta Phys. Polon. }{\bf #1~} #3 (19#2)}
\def\rnc#1#2#3{{\it Riv. Nuovo Cim. }{\bf #1~} #3 (19#2)}
\def\ap#1#2#3{{\it Ann. Phys. }{\bf #1~} #3 (19#2)}
\def\ptp#1#2#3{{\it Prog. Theor. Phys. }{\bf #1~} #3 (19#2)}
\def\spu#1#2#3{{\it Sov. Phys. Usp.}{\bf #1~} #3 (19#2)}
\def\apj#1#2#3{{\it Ap. J.}{\bf #1~} #3 (19#2)}
\def\epj#1#2#3{{\it Eur.\ Phys.\ J. }{\bf C#1~} #3 (19#2)}
\def\pu#1#2#3{{\it Phys.-Usp. }{\bf #1~} #3 (19#2)}

\end{document}